# Automated Atlas-based Segmentation of Single Coronal Mouse Brain Slices using Linear 2D-2D Registration

Sébastien Piluso[1,2], Nicolas Souedet[1], Caroline Jan[1],
Cédric Clouchoux[2], Thierry Delzescaux[1]

*Abstract*—A significant challenge for brain histological data analysis is to precisely identify anatomical regions in order to perform accurate local quantifications and evaluate therapeutic solutions. Usually, this task is performed manually, becoming therefore tedious and subjective. Another option is to use automatic or semi-automatic methods, among which segmentation using digital atlases co-registration. However, most available atlases are 3D, whereas digitized histological data are 2D. Methods to perform such 2D-3D segmentation from an atlas are required. This paper proposes a strategy to automatically and accurately segment single 2D coronal slices within a 3D volume of atlas, using linear registration. We validated its robustness and performance using an exploratory approach at whole-brain scale.

## I. INTRODUCTION

Over the last decades, usage of histological staining techniques coupled to brain microscopic imaging has dramatically increased in neuroscience and provides new tools to study the brain. One major challenge in brain histology is to precisely locate anatomical regions to perform accurate quantification of tissue staining. Digital 3D brain anatomical atlases have played a significant role in segmenting brain data, and have been widely used with different modalities, especially in preclinical research [1]. Despite those breakthroughs, brain parcellation is still mostly performed manually by neurobiologists, making this task tedious and non-reproducible (inter- and intra-operator bias).

Previous studies have already validated the automatic segmentation of reconstructed histological volumes within 3D atlas [1]. Depending on the histological protocol, the number of produced sections can dramatically change. In some cases, a large number of sections makes 3D reconstruction possible and enables the use of such 3D-3D registration strategies with atlases. However, most biological studies rely on a few slices only, thus characterized as *single* slices, precluding 3D reconstruction. In this case, neurobiologists manually perform parcellation relying on anatomical region correspondences, or visually identify the slice position within the 3D atlas to further perform 2D-2D registration between the identified atlas slice and the slice to be segmented. However, manual processes lack reliability [5] and possible 3D tilt in orientation can occur after brain positioning before the cutting process. Depending on the magnitude of those tilting angles, 2D-2D coronal anatomical correspondence with digital atlas may be compromised.

Automated methods are then needed to perform atlas segmentation of single 2D histological slices. As most histological mouse brain studies are performed in the coronal incidence, dedicated state-of-the-art related automated tools have been developed. In addition to 2D-2D registration (rigid, affine, elastic), most researches focused on three parameters: the position of the single coronal slice along the antero-posterior (AP) axis (*z-position*), and two angles around the infero-superior axis ($\varphi$) and left-right axis ($\theta$), qualifying the tilt in orientation [2][4][5].

To date, a few tools have been proposed, using 2D-2D registration coupled to a metric estimating the similarity between two anatomical 2D images. A first study [2] provided an automated way to estimate the two angles $\varphi$ and $\theta$ using the histogram of oriented gradient (HOG) similarity metric [3], given a manually predefined *z-position*. Another study proposed *QuickNII* [4], a semi-automatic software that allows to manually perform linear deformations on images to estimate the best matching plane within a 3D atlas. The method includes the possibility to explore different $\varphi$ and $\theta$ angles. A recent study presented *AMaSiNe* [5], an automatic method to estimate both the *z-position* of the slice(s) to be segmented, as well as $\varphi$ and $\theta$ angles, using a HOG similarity metric based on feature points soundly chosen on images. Whilst some researches have been carried out on the topic, most of the methods still require manual intervention. Only one study has attempted to investigate the automatic detection of the *z-position* of single slices [5], however lacking reproducibility in estimating the z-position, $\varphi$ and $\theta$ angles for one single slice.

[1]CEA, Fundamental Research Division (DRF), Institute Jacob, Molecular Imaging Research Center (MIRCen), CEA Fontenay-aux-Roses, France.
[2]WITSEE, Paris, France.

Our paper proposes a new fully-automated method to precisely determine the *z-position* along the AP axis of a single 2D experimental coronal slice within a 3D anatomical atlas volume, when no tilting $\varphi$ and $\theta$ angles are introduced by the slicing task. Linear 2D-2D registration is thereafter performed between the experimental single slice and its corresponding section in the atlas, determined by the method. We independently made used this strategy for each section of a whole experimental mouse brain to evaluate its performance at whole-brain scale. At last, we separately induced significant $\varphi$ and $\theta$ rotations to the experimental data in order to quantify their respective influence on the results.

## II. MATERIAL AND METHODS

### A. Mouse anatomical brain dataset

In this paper, we used the left hemisphere of the template from the Allen Mouse Brain Atlas [6], one of the most popular mouse atlas in neurobiology. This template was constructed as an average autofluorescence of 1,675 serial two-photons tomography C57/Bl6J mouse brains. In this work, we aimed at segmenting 2D single slices from the autofluorescence of an experimental clarified half mouse brain imaged using light sheet microscopy [7].

On the one hand, we considered each single 2D coronal slice $s_e$ to be independently segmented, among the 417 slices extracted from the experimental 3D autofluorescence volume. On the other hand, we considered a succession of 436 2D adjacent slices $s_t$, extracted from the atlas template volume in the same incidence. A preprocessing stage consisted in resampling all slices at an isotropic resolution of 25 µm and their field of view was adjusted. This resolution enabled to register one slice to another efficiently (preserving enough details) while being performed in less than one minute on a single core [7].

We evaluated $\varphi$ and $\theta$ angles between the experimental volume and the template using 3D-3D rigid registration [1]. Regarding the data we used in this paper, both angles were considered as negligible (< 1°), validating the choice of a 2D-2D registration approach to focus on single slice AP position estimation.

In order to estimate the impact of rotations, we generated two additional sets of experimental brain data from the autofluorescence volume, with simulated tilting angles between both volumes: (1) for $\theta = 10°$, (2) for $\varphi = 10°$.

### B. Registration method for a single slice and application at whole-brain scale

The general purpose of the method is to find the most similar $s_t$ slice among the template sections for each experimental single slice $s_e$, using linear registration. This work relies on two registration approaches: rigid and affine, based on Block-Matching techniques using correlation coefficient as a similarity criterion [8]. After registration, an independent global slice-to-slice similarity between the experimental slice $s_e$ and every registered template slice $s_t$ was evaluated using normalized mutual information (NMI) [9]. This metric, based on contrast difference within images, has demonstrated its robustness for calculating similarity between images from different modalities. As rigid and affine registrations have different degrees of freedom (translation and rotation for rigid, and scaling and shearing for affine), they both contribute to optimal registration. As a third strategy, a hybrid approach consisted in calculating the mean of rigid and affine NMI scores. The optimal template slice candidate $\hat{s}_t$ was then determined by detecting the maximal NMI value in the corresponding vector for each strategy.

In order to evaluate the method performance at whole-brain scale, it was tested for each $s_e$ slice of the 417 slices composing the autofluorescence volume, for the three approaches. All NMI vectors for each $s_e$ slice were concatenated into two NMI cartographies (rigid and affine), the third approach being the mean of those two cartographies.

Similar cartographies were calculated for the two datasets including tilting angles, for $\theta = 10°$ and $\varphi = 10°$ respectively.

### C. Validation of the method

In order to evaluate the relevance of the best $\hat{s}_t$ slice candidate estimated from each approach, a reference vector of correspondences was manually defined. A neurobiologist was asked to manually identify the optimal corresponding $\check{s}_t$ slice, for 30 $s_e$ slices regularly distributed over the entire experimental volume, resulting in as many vectors of paired values ($s_e$, $\check{s}_t$). A linear regression was calculated between the 30 $s_e$ and the 30 $\check{s}_t$ slice numbers, giving a linear relation $f(s_e) = a.s_t + b$ for all $s_e$ slices from the experimental volume. This regression, qualified by its $R^2$ coefficient of determination, was then considered as the ground truth for the determination of the $\check{s}_t$ corresponding best slice for each $s_e$ slice. Indeed, as both volumes contain same progressive proportional anatomies, a linear relationship between $s_e$ and $\hat{s}_t$ slices was assumed (confirmed by the ground truth). Thus, linear regressions were evaluated considering all paired values ($s_e$, $\hat{s}_t$) from each strategy, characterized by their $R^2$ value.

Performance was evaluated using $\Delta_{sn} = |\check{s}_t - \hat{s}_t|$ (absolute difference between the estimated and the expert slice numbers), where $\Delta_{sn} = 0$ corresponds to the same z-position estimation as experts.

Slice-to-slice registration quality control was assessed using non-weighted mean Dice score on regions of interest of registered images. A neurobiologist manually delineated six major anatomical regions of different sizes (*cortex*, *striatum*, *thalamus*, *hippocampus*, *substantia nigra* and *globus pallidus*) on 5 coronal slices of interest chosen from the experimental volume, as well as on their corresponding registered $\hat{s}_t$ sections determined from the three approaches.

At last, Dice score and $R^2$ were calculated for each of the two cartographies including tilting angles, to evaluate their influence on the results. In these cases, the $\Delta_{sn}$ criterion was not calculated, the rotations making it unrealistic to define one single corresponding coronal slice in the atlas template.

### D. Implementation details

Considering the large amount of calculations, the pipeline of the method was run using distributed computing on multiple microprocessors using *SomaWorkflow* library of BrainVISA software [10]. This work was conducted on a workstation with Ubuntu 16.04 LTS 64-bits on Intel® Xeon(R) CPU E5-2620 v2 @ 2.10GHz × 24 (24 computing cores) and 128 GB of Random Access Memory (RAM).

## III. RESULTS

The three vectors corresponding to each approach for a sample slice ($s_e = 240$) are presented on Fig. 1. Maximum NMI value detection for each method gave different $\hat{s}_t$ candidate slices, where the closest to the expert was found for the mean strategy ($\Delta_{sn} = 4$, lowest value). Evaluation of $\Delta_{sn}$ extended to the entire $s_e$ slices set and $R^2$ values are presented in Tab. I. Maximum detection on the mean NMI vector showed best scores (lower $\Delta_{sn}$) on average.

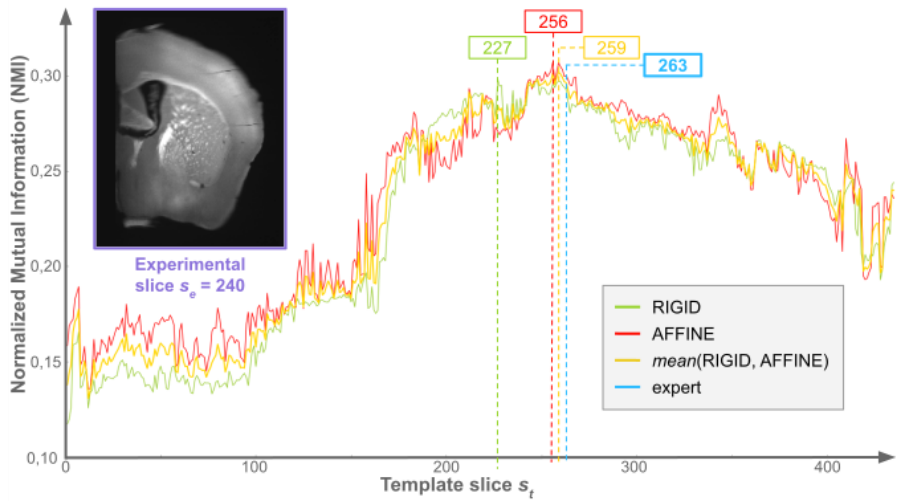

**Fig. 1.** NMI vectors calculated between the coronal slice $s_e = 240$ and each $s_t$ template slice using three approaches (rigid in green, affine in red, rigid and affine mean in yellow), as well as the ground truth corresponding to the most similar $s_t$ slice determined by the expert (in blue).

**TABLE I.** Evaluation of the performance of the coronal $\hat{s}_t$ best candidate estimation method ($\Delta_{sn}$ shift in number of slices, and $R^2$ coefficient of determination) for the three approaches at whole-brain scale, as well as the registration quality (mean Dice) in 5 slices, compared with the expert.

| CORONAL | RIGID | AFFINE | MEAN | EXPERT |
|---|---|---|---|---|
| mean($\Delta_{sn}$) | 10.8 ± 35.4 | 7.0 ± 12.1 | 4.0 ± 4.8 | - |
| $R^2$ | 0.88 | 0.98 | 0.99 | 0.99 |
| mean(**Dice**) | 0.89 ± 0.06 | 0.90 ± 0.06 | 0.90 ± 0.06 | 0.88 ± 0.08 |

According to the tested slice $s_e$, rigid and affine vectors gave each in turn closest results compared to expert pairing. The hybrid approach performed the best $\hat{s}_t$ candidate estimation at the entire brain scale. The mean approach evaluated any coronal slice location within a mean precision of ~100 µm (~4 slices) along the AP axis (see Tab. I).

Rigid and affine NMI cartographies were merged in Fig. 2, along with the ground truth. Each cartography included more than 180,000 co-registrations, computed in parallel on 20 cores and representing about 130 hours (~53 seconds per job).

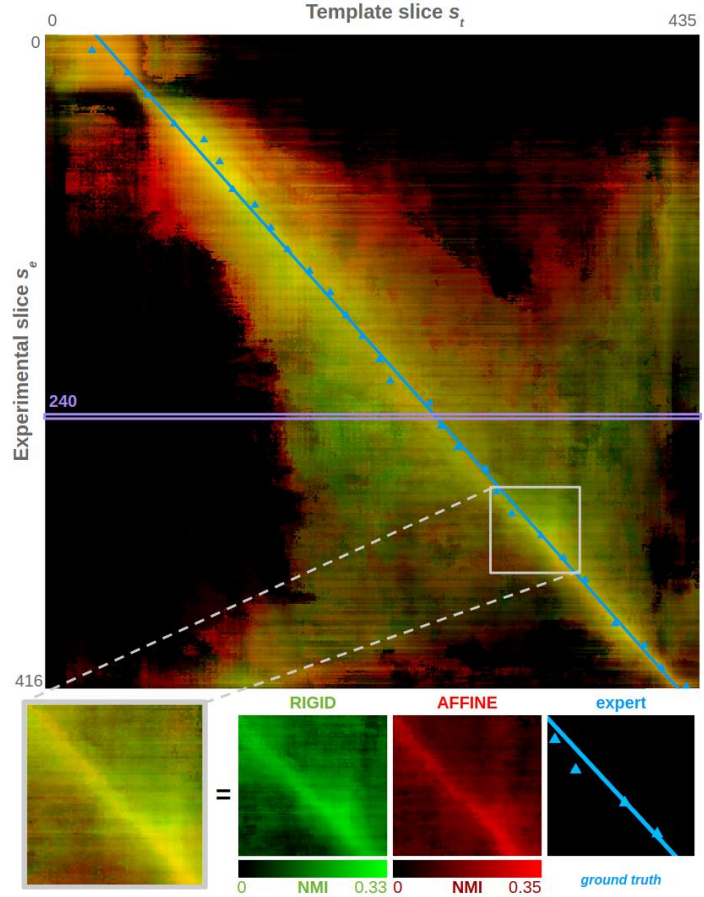

**Fig. 2.** Cartography of similarity (NMI values) between each $s_e$ (vertical) and each $s_t$ (horizontal) coronal slices for rigid (green) and affine (red) registration, as well as the expert linear regression (blue) superimposed. Different contributions are splitted for a given example of a grey square area. Yellow coloration corresponds to areas where both approaches (rigid and affine) have together highest NMI scores.

**TABLE II.** Evaluation of the performance of the coronal $\hat{s}_t$ best candidate estimation method ($R^2$ coefficient of determination) for the three approaches, as well as the registration quality (mean Dice) for datasets including respective tilting angles $\theta = 10°$ and $\varphi = 10°$.

| $\theta = 10°$ | RIGID | AFFINE | MEAN |
|---|---|---|---|
| $R^2$ | 0.96 | 0.79 | 0.97 |
| mean(**Dice**) | 0.67 ± 0.33 | 0.68 ± 0.35 | 0.74 ± 0.31 |
| $\varphi = 10°$ | RIGID | AFFINE | MEAN |
| $R^2$ | 0.48 | 0.94 | 0.80 |
| mean(**Dice**) | 0.38 ± 0.40 | 0.71 ± 0.28 | 0.56 ± 0.40 |

Fig. 2 shows a global diagonal shape where best NMI scores were found, confirming the linear relationship between the slices of the two anatomical volumes which are geometrically close. In contrast, the two extremity parts of the cartography present specific results: the top left part does not highlight clear maximal NMI value, whilst the last 15 $s_e$ lines highlight 2 maximal NMI peak values. However, the three approaches demonstrated a strong NMI linearity relationship between $s_e$ and $s_t$ slices, with high $R^2$ values (see Tab. I). In particular, affine and hybrid strategies showed $R^2$ values similar to expert evaluation (> 0.97). Moreover, averaged Dice scores evaluated from expert manual segmentations showed similar high success rates (≥ 0.88) for all approaches (Tab. I).

Results from datasets including tilting angles present a lower mean overlap between segmented regions (Dice < 0.75 whatever the strategy), with a rather high variability (Dice standard deviation ≥ 0.28) (Tab. II). Some regions fit well, whereas others are not part of the $\hat{s}_t$ segmented slice. Moreover, $\varphi$ tilting orientation seems to give even worse Dice scores than $\theta$ does. $R^2$ coefficients of determination also show a high variability at whole-brain scale according to the strategy, and sometimes are very low (< 0.50).

## IV. DISCUSSION

In this paper, we proposed an automated method to estimate the AP position of a single coronal slice, when no tilting angles is introduced. This method is based on an exhaustive exploratory approach relying on linear 2D-2D registration (rigid and affine) between the experimental slice and all template atlas slices. Registration quality was assessed using similarity estimation, and the method has been tested at whole-brain level. The hybrid approach, based on the mean calculation of both rigid and affine registration, performed a precise estimation of the *z-position* for any coronal experimental slice, resulting in segmentations showing equivalent dice to experts for the example slices we assessed.

Rigid registration was relevant to identify the geometrically closest slice to the experimental one. More specifically, the additional degrees of freedom of the affine registration (scaling, shearing) allowed for more precise transformations. However, NMI combined cartography (Fig. 2) showed that a good initialization of *z-positions* for the best slice candidates was required for affine registration, otherwise leading to skewed slice identification (large transformations should be forbidden to avoid excessive and unrealistic deformations of tissues).

A visual analysis of the cartography highlighted regions with selective prevalence of rigid and/or affine registration(s), probably due to anatomical content of the slices (Fig. 2). As mentioned in the Results section, the first top left square of the cartography does not clearly emphasize maximum NMI values corresponding to potential $\hat{s}_t$ slice candidate for the first 40 $s_e$ slices. This region of the cartography corresponds to the *olfactory bulb* only, which is a rounded shape containing no specific other contrast. This could be one of the reasons why most of the $s_t$ slices containing only the *olfactory bulb* globally match those $s_e$ slices, independently from applied transformation. Similar reasons could explain the two maximum peak values for the last 15 $s_e$ slices: global brain proportions in the $s_t$ slice images corresponding to those peaks are the same as for the $s_e$ slices. This difference is noticeable especially for rigid registration, where wrong maximal peak values are detected by the method and result in a misidentification of the $\hat{s}_t$ slice candidates for the last 5 $s_e$ slices. This is why $\Delta_{sn}$ presents a high variability (standard deviation > 35 slices in average for the rigid registration).

Validation procedures of the different approaches were performed at two levels: 1) evaluation of the relevance of the selected atlas slice compared to the expert identification (based on $\Delta_{sn}$ metric), and 2) evaluation of the relevance of resulting segmentation using registration (based on Dice coefficient). The first allowed a reliable ranking of the various approaches. However, the Dice coefficient was very high in all cases on anatomical regions of various sizes even when the candidate slice of the atlas was not optimal. This criterion provided an essential indicator for the segmentation quality evaluation, but is not a reliable criterion to evaluate the choice of the slices.

The preliminary evaluation of the robustness to tilting demonstrated limitations: a 10° angle magnitude significantly affected the final segmentation and led to a misidentification of a *z-position* for a single $s_e$ slice within the atlas template. Moreover, even if the *z-position* estimation is roughly correct, global shape, size and position of some regions are still affected by the added simulated rotation (as shown by Dice scores in Tab. 2). Segmentation of small regions (*globus pallidus, substantia nigra* for example) were more affected than larger regions (*cortex* for instance). These results suggest that the proposed method is suitable in a relatively strict 2D-2D registration context, however providing good segmentation results in case of slight tilting angles but limited to large anatomical regions. However, it is noteworthy that 3D tilting introduced during the cutting process is often very limited, and usually smaller than the simulated values used in our experiment. Therefore, more tests are needed to evaluate the influence of tilting on the proposed method by generating new datasets with a larger range of angles, as well as combining them together.

## V. Conclusion

Our study presented a robust automated 2D-2D single coronal slice atlas segmentation method, assessing its performance using an exploratory registration approach across the whole mouse brain anatomy on histological data. After having investigated anatomical similarities between 3D coherent volumes that defined our first study approach, same strategies will be applied using other sets of data, such as block-face imaging, or cresyl staining, presenting geometric distortions. Moreover, attempts will be conducted to deploy the codes on high performance calculation infrastructures[1] in order to test the method on larger monkeys brain data.

Future work will also consider the possibility to take into account $\varphi$ and $\theta$ tilting angles to define an optimal oblique plane to perform proper anatomical segmentation with the atlas volume.

## Compliance With Ethical Standards

The experimental procedures involving animal models described in this paper were approved by the Institutional Animal Care and Ethics Committee.

## Acknowledgment

[1]This work was granted access to the HPC resources of TGCC under the allocation 2019-(A0040310374) made by the GENCI.